\documentclass[12pt]{book}

\usepackage[dvips]{graphicx,color}
\usepackage{makeidx,high}
\makeauthorindex
\makeindex

\begin{document}

\BookTitle{\itshape Extremely High Energy Cosmic Rays}
\CopyRight{\copyright 2002 by Universal Academy Press, Inc.}
\pagenumbering{arabic}

\chapter{Signatures of Protons in UHECR}

\author{V. Berezinsky\\
{\it INFN, Laboratori Nazionali del Gran Sasso, I-67010 Assergi (AQ), Italy}\\
A. Gazizov\\
{\it DESY Zeuthen, Platanenallee 6 D-15738 Zeuthen, Germany}\\
S. Grigorieva\\
Institute for Nuclear Research, 60th October Revolution Prospect 7A, 
117312 Moscow, Russia}

\AuthorContents{V.\ Berezinsky, A.\ Gazizov and S.\ Grigorieva} 
\AuthorIndex{Berezinsky}{V.} 
\AuthorIndex{Gazizov}{A.} 
\AuthorIndex{Grigorieva}{S.}

\section*{Abstract}
We demonstrate that the energy spectra of Ultra High Energy Cosmic
Rays (UHECR) as observed by AGASA, Fly's Eye, HiRes  and Yakutsk 
detectors, have the imprints of UHE proton interaction with the CMB 
radiation as the dip centered at $E\sim 1\times 10^{19}$~ eV,
beginning of the GZK cutoff, and very good agreement with
calculated spectrum shape. This conclusion about proton composition 
agrees with recent HiRes data on elongation rate that support the 
proton composition at $E\geq 1\times 10^{18}$~eV. The visible bump 
in the spectrum at $E \sim 4\times 10^{19}$~eV is not caused by pile-up 
protons, but is an artifact of multiplying the spectrum by $E^3$.
We argue that these data, combined
with small-angle clustering and correlation with AGN (BL Lacs),    
point to the AGN model of UHECR origin at energies 
$E \leq 1\times 10^{20}$~eV. The events at higher energies and the excess 
of the events at $E \geq 1\times 10^{20}$~eV , which is observed by 
AGASA (but absent in the HiRes data) must be explained by another component
of UHECR, e.g. by UHECR from superheavy dark matter.
\section{Introduction}
The nature of signal carriers of UHECR is not yet 
established. The most 
natural primary particles are extragalactic protons.
Due to interaction with the CMB radiation the UHE protons
from extragalactic sources are predicted to have a sharp
steepening of energy spectrum, so called GZK cutoff (Greisen 1966,
Zatsepin, Kuzmin 1966).
For uniformly distributed sources, the GZK
cutoff is characterized by energy $E_{1/2}$, where the integral
spectrum calculated with energy losses taken into account becomes
twice lower than the power-law extrapolation from low energies
(Berezinsky, Grigorieva 1988); $E_{1/2} = 5.7\times 10^{19}$~eV.

\begin{figure}[htb]
  \begin{center}
    \includegraphics[height=20pc]{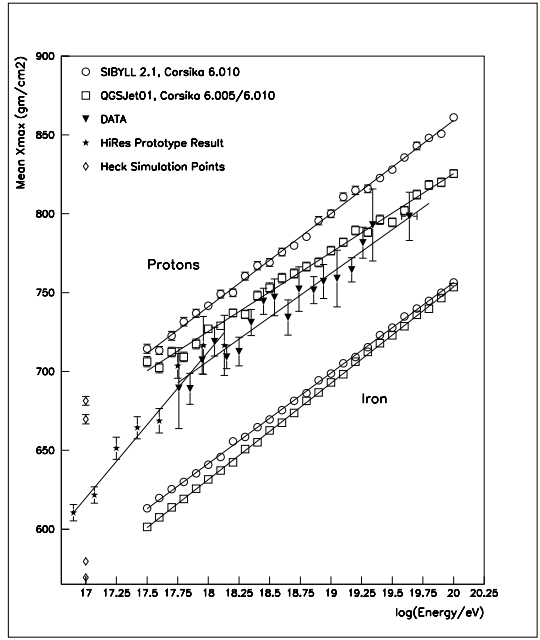}  
\end{center}
  \caption{The HiRes data (Sokolsky 2002) on mass
composition (preliminary). The measured $x_{\rm max}$ at 
$E\geq 1\times 10^{18}$~eV (triangles) are in a good agreement with 
QGSJet-Corsika prediction for protons.}
  \label{hires}
\end{figure}

There are two other signatures of extragalactic protons in the
spectrum: dip and bump (Hill and Schramm 1985, Berezinsky and
Grigorieva 1988, Yoshida and Teshima 1993, Stanev et al 2000). 
The dip is produced due  
$p+\gamma_{\rm CMB} \to p+e^++e^-$ interaction at energy 
$E \sim 1\times 10^{19}$~eV. The bump is produced by pile-up protons  
which loose energy in the GZK cutoff. As it was demonstrated by 
Berezinsky and Grigorieva (1988) (see also Stanev et al 2000), the
bump is clearly seen from a single source at large redshift $z$, but
it practically disappears in the diffuse spectrum, because individual 
peaks are located at different energies. We shall demonstrate here
that what is seen now in the observed spectrum as a broad bump is 
an artifact caused by multiplication of the spectrum to $E^3$.

As we shall demonstrate here, the observed spectra by AGASA, Fly's Eye,
HiRes and Yakutsk arrays  have the imprints  of extragalactic protons 
in the form of the dip centered at $E \sim 1\times 10^{19}$~eV and 
of the beginning of the GZK cutoff, and in the form of good agreement
between predicted and observed spectra. The measurement of the
atmospheric height 
of EAS maximum, $x_{\rm max}$, in the HiRes experiment (for preliminary data
see Fig.\ref{hires}) gives the strong evidence in favor of pure proton
composition at $E \geq 1\times 10^{18}$~eV. Yakutsk data also favor 
the proton composition at $E\geq 1\times 10^{18}$~eV (Glushkov et al 2000). 

At what energy the extragalactic component sets in?

\begin{figure}[htb]
  \begin{center}
    \includegraphics[height=6cm]{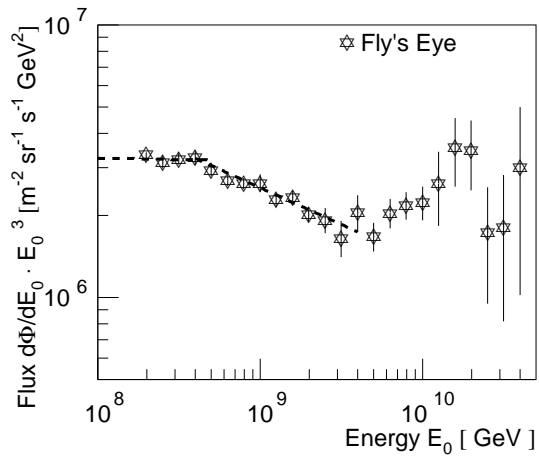}
  \end{center}
 \caption{The second knee from Fly's Eye data (Bird et al 1994).}
  \label{FE}
\end{figure}

According to the KASCADE data (Kampert 2001,Hoerandel 2002 ), the spectrum of galactic 
protons has a steepening at $E\approx 3\times 10^{15}$~eV (the first
knee), helium nuclei - at $E \approx 6\times 10^{15}$~eV, and carbon
nuclei - at $E \approx 1.5\times 10^{16}$~eV. It confirms the 
rigidity-dependent
confinement with critical rigidity $R_c=E_c/Z \approx 3\times 10^{15}$~eV. 
Then galactic iron nuclei are expected to have the critical energy of
confinement at $E_c \sim 1\times 10^{17}$~eV, and extragalactic
protons can naturally dominate at $E \geq 1\times 10^{18}$~eV. This
energy is close to the energy of the second knee (Akeno - 
$6\times 10^{17}$~eV, Fly's Eye - $4\times 10^{17}$~eV, HiRes -  
$7\times 10^{17}$~eV and Yakutsk - $8\times 10^{17}$~eV). In Fig.\ref{FE}
the second knee is shown according to Fly's Eye observations. It illustrates
a possible transition to extragalactic cosmic rays. 

The AGASA data (Hayashida et al 1999) 
which shows an excess of events from regions of Galactic Center and Cygnus  
at $ E\sim 10^{18}$~eV further confirm the picture outlined
above. Indeed, unconfined galactic particles (most notably protons) 
propagate quasi-rectilinearly and show the direction to the sources,
while the diffuse flux is dominated by extragalactic protons. 

The model of galactic cosmic rays developed by Biermann et al (2003)
also predicts the second knee as the ``end'' of galactic cosmic rays
(iron nuclei) due to rigidity bending in wind-shell around SN. 
The extragalactic component became the dominant one at energy 
$E \sim 1\times 10^{18}$~eV (see Fig.1 in Biermann et al 2003).  

The good candidates for the  sources of observed UHE protons are AGN. 
They can accelerate protons up to energy $E_{\rm max} \sim 10^{21}$~eV
(Biermann and Streitmatter 1987, Ipp and Axford 1991, Rachen and
Biermann 1993), they have power to provide the observed flux of UHE 
protons (Berezinsky, Gazizov and Grigorieva 2002) and 
finally there is direct correlation (Tinyakov and Tkachev 2001, 2003
and references therein) 
between directions of arrival of 
UHE particles with energies $(4 - 8)\times 10^{19}$~eV and directions
to BL Lacs, which comprise some particular class of AGN. 

Does it mean that UHECR puzzle has been already resolved in most 
conservative way? 

In this model AGN cannot be the sources of observed particles with energy
$E \geq 1\times 10^{20}$~eV (Berezinsky, Gazizov, Grigorieva 2002): 
the attenuation length for a proton of this energy is smaller than 135 Mpc, 
and correlation with AGN would be seen for all these particles,
contrary to observations. The particles observed at 
$E \geq 1\times 10^{20}$~eV, in particular those detected in AGASA,
imply the presence of another component, e.g. produced by decays of 
superheavy DM.

\section{Calculation of the spectra: dip and GZK cutoff}
We calculate the UHE proton spectra in the model with uniform
distribution of the sources, with CR luminosity of a source $L_p(z)$,
and CR emissivity ${\cal L}(z)=n(z)L_p(z)={\cal L}_0(1+z)^m$, where 
$n(z)$ is comoving space density of the sources at epoch with redshift $z$, 
index 0 refers to $z=0$, and $m$ describes cosmological evolution of
the sources. We assume the generation spectrum of a source
\begin{equation}
Q_g(E_g,z)=\frac{L_p(z)}{\ln\frac{E_c}{E_{\rm min}}+\frac{1}{\gamma_g-2}}
q_{\rm gen}(E_g),
\label{gen}
\end{equation}
\begin{equation}
q_{\rm gen}(E_g)=\left\{ \begin{array}{ll}
1/E_g^2                      ~ &{\rm at}~~ E_g \leq E_c\\
E_c^{-2}(E_g/E_c)^{-\gamma_g}~ &{\rm at}~~ E_g \geq E_c
\end{array}
\right.
\label{q-gen}
\end{equation}

The diffuse spectrum can be calculated as 
\begin{equation}
J_p(E)=\frac{c}{4\pi}~\frac{{\cal L}_0 } 
{\ln\frac{E_c}{E_{min}}+\frac{1}{\gamma_g-2}} \int_0^{z_{max}}
dt(1+z)^m q_{\rm gen}\left(E_g(E,z),E\right) \frac{dE_g}{dE},
\label{diff}
\end{equation}
where $E_g(E,z)$ is generation energy of a proton at epoch $z$, $dE_g/dE$
is given in Refs. (Berezinsky, Grigorieva 1988, and Berezinsky,
Gazizov,Grigorieva 2002), and $dt$ is given as  
\begin{equation}
dt=\frac{dz}{H_0(1+z)\sqrt{\Omega_m(1+z)^3+\Omega_{\Lambda}}},
\label{dt/dz}
\end{equation}
with $H_0, \Omega_m,$ and $\Omega_{\Lambda}$ being the Hubble
constant, relative cosmological density of matter and vacuum energy,
respectively. 
\begin{figure}[htb]
  \begin{center}
    \includegraphics[height=22pc]{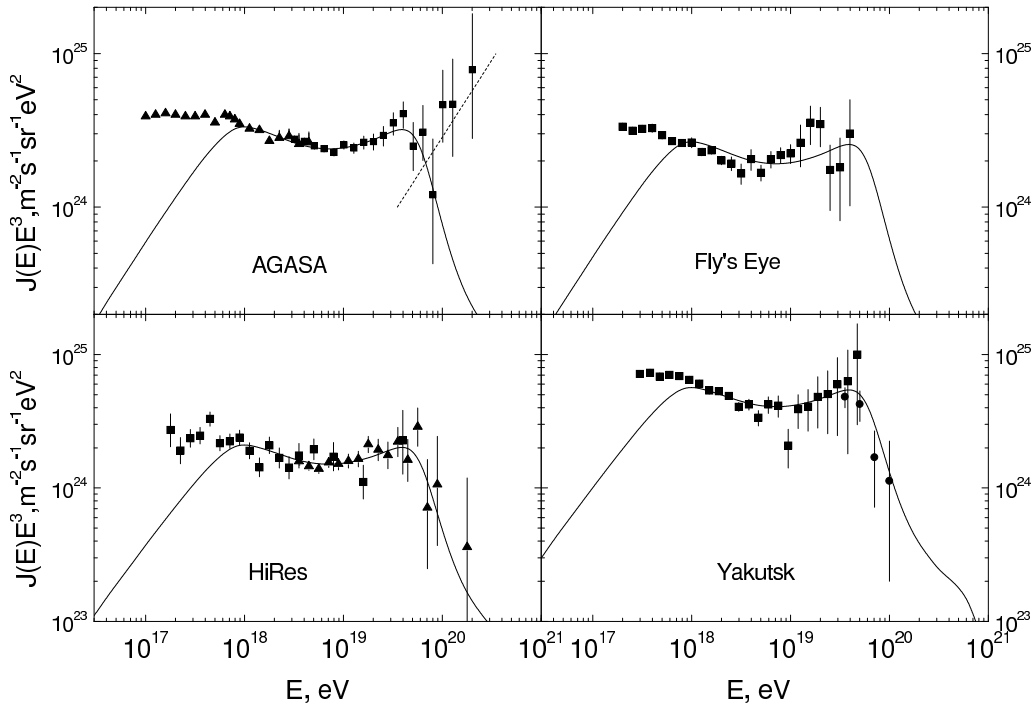}
  \end{center}
 \caption{Comparison of calculated spectra for non-evolutionary model
with experimental data. The normalization has been first performed by
the AGASA data at $E=1\times 10^{18}$~eV, adjusting the emissivity 
${\cal L}_0$. To fit the data of HiRes, Fly's Eye and Yakutsk the emissivity
${\cal L}_0$ has been scaled by factors 0.63, 0.8 and 1.7, respectively.}
\label{spectra}
\end{figure} 
In Fig.\ref{spectra} we present the calculated spectra compared with 
AGASA (Takeda et al 2002), HiRes (Abu-Zayyad et al 2002a), 
Fly's Eye (Bird et al 1994) and Yakutsk (Glushkov and Pravdin 2001) data. 
The dip is well
confirmed by all data. HiRes and Yakutsk data agree well with
existence of GZK cutoff, while the AGASA data show the excess of
events at $E\geq 1\times 10^{20}$~eV.  

For the curves above we used the following parameters: 
$\gamma_g= 2.7$,~ $E_c=1\times 10^{18}$~eV, $E_{\rm max}=1\times 10^{21}$~eV,
$H_0=70$~km/s Mpc, 
$\Omega_m=0.3$,~ $\Omega_{\Lambda}=0.7$. We have chosen the CR emissivity 
${\cal L}_0=3.5\times 10^{46}$~erg/Mpc$^3$yr  
to normalize the AGASA
data, and scaled ${\cal L}_0$ by factors 0.63,~ 0.80 and 1.7 to
normalize the data of HiRes, Fly's Eye and Yakutsk, respectively. 

\section{$E_{1/2}$ as characteristic of the GZK cutoff}
\begin{figure}[htb]
 \begin{center}
    \includegraphics[height=18pc]{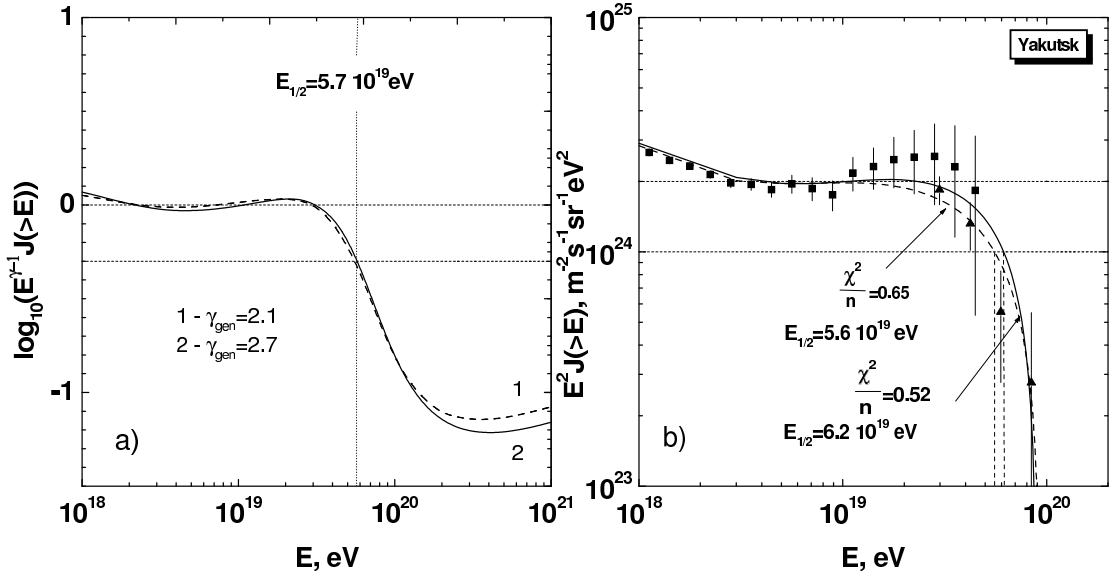}
  \end{center}
 \caption{$E_{1/2}$ as numerical characteristic of the GZK cutoff.        
In panel a) the calculations for different $\gamma_{\rm gen}$ are presented.
In panel b) $E_{1/2}$ is found from the integral spectrum of the
Yakutsk array using two fits of the integral spectrum.}
  \label{E_1/2}
\end{figure}

$E_{1/2}$ is the energy where the flux calculated with energy losses
becomes twice less than power-law extrapolation of integral spectrum.  
In Fig.\ref{E_1/2}a the function $E^{(\gamma-1)}J(>E)$ is plotted as
function of energy ($\gamma>\gamma_g$ is the effective index). For
wide range of generation indices $2.1 \leq \gamma_g \leq 2.7$ the
cutoff energy is the same, $E_{1/2} \approx 5.7\times 10^{19}$~eV.
We have determined $E_{1/2}$ from the Yakutsk data. For this we found two
fits  of the Yakutsk integral spectrum with help of trial functions,
as shown in Fig.\ref{E_1/2}b. They have good $\chi^2/n$ equal to 0.65
and 0.52. The corresponding values of $E_{1/2}$, ~$5.6\times 10^{19}$~eV  
and $6.2\times 10^{19}$~eV, agree well with the theoretical value.
Note, that in the fits above $\chi^2/n$ are the formal values from which 
probabilities cannot be calculated in the standard way, because the points
in the integral spectrum are correlated quantities. 

This analysis obviously cannot be extended to the AGASA integral
spectrum, because  of too many events at the highest
energies. Unfortunately, we do not have the HiRes integral spectrum
to perform the analysis as that above. 

\section{Bump in the diffuse spectrum}

\begin{figure}[htb]
 \begin{center}
    \includegraphics[height=23pc]{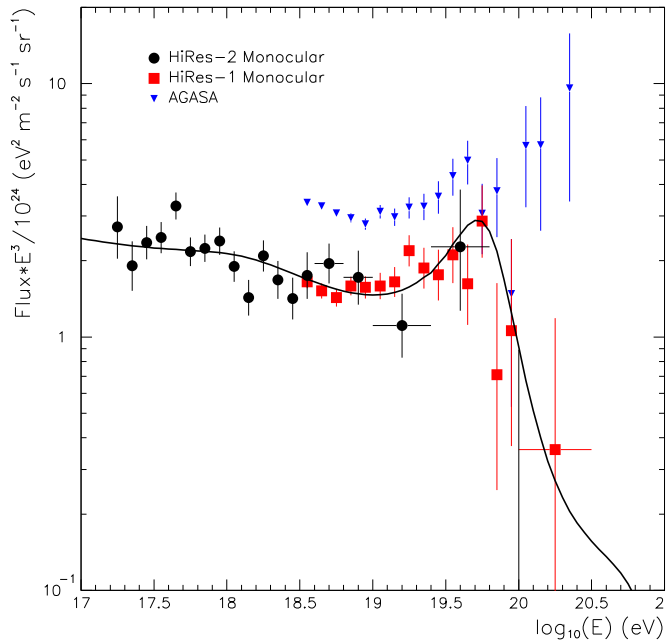}
  \end{center}
 \caption{HiRes spectrum from Abu-Zayyad et al 2002b.   
} 
  \label{HRbump}
 \end{figure}

The bump is distinctly  seen in the measured spectra when they are  
multiplied to $E^3$ (as example see the HiRes spectrum in Fig.\ref{HRbump}). 

Is this bump really composed by pile-up protons?

To discuss the bump produced by pile-up protons it is convenient to 
introduce the  {\em modification factor} $\eta$  
(Berezinsky and Grigorieva 1988), defined 
for the power-law generation spectrum according to 
Eqs.(\ref{gen}) - (\ref{diff}).
\newpage
Without loss of generality, one can
assume the power-law generation spectrum, and present the diffuse
spectrum $J_p(E)$ as the product of unmodified flux, $J_{\rm unm}$,
and modification factor $\eta(E,z_{\rm max})$, which describes the distortion
of spectrum by energy losses:
\begin{equation}
J_{\rm unm}(E)=\frac{c}{4\pi}(\gamma_g-2)\frac{{\cal L}_0}{H_0}E^{-\gamma_g},
\end{equation}
and
\begin{equation}
\eta(E,z_{\rm max})=\int_0^{z_{\rm max}}dz \frac{\lambda^{-\gamma_g}(E,z)}
{(1+z)\sqrt{(1+z)^3\Omega_m+\Omega_{\Lambda}}}\frac{dE_g}{dE},
\label{mfactor}
\end{equation}
where $\lambda(E,z)=E_g(E,z)/E$. When $E$ is small, $\lambda=(1+z)$,~
$dE_g/dE=(1+z)$ and modidification factor tends to
energy-independent value
\begin{equation}
\eta(z_{\rm max})= \int_0^{z_{\rm max}}dz\frac{(1+z)^{-\gamma_g}}
{\sqrt{(1+z)^3\Omega_m+\Omega_{\Lambda}}}.
\label{eta-lim}
\end{equation}
\newpage
\begin{figure}[ht]
 \begin{center}
    \includegraphics[height=25pc]{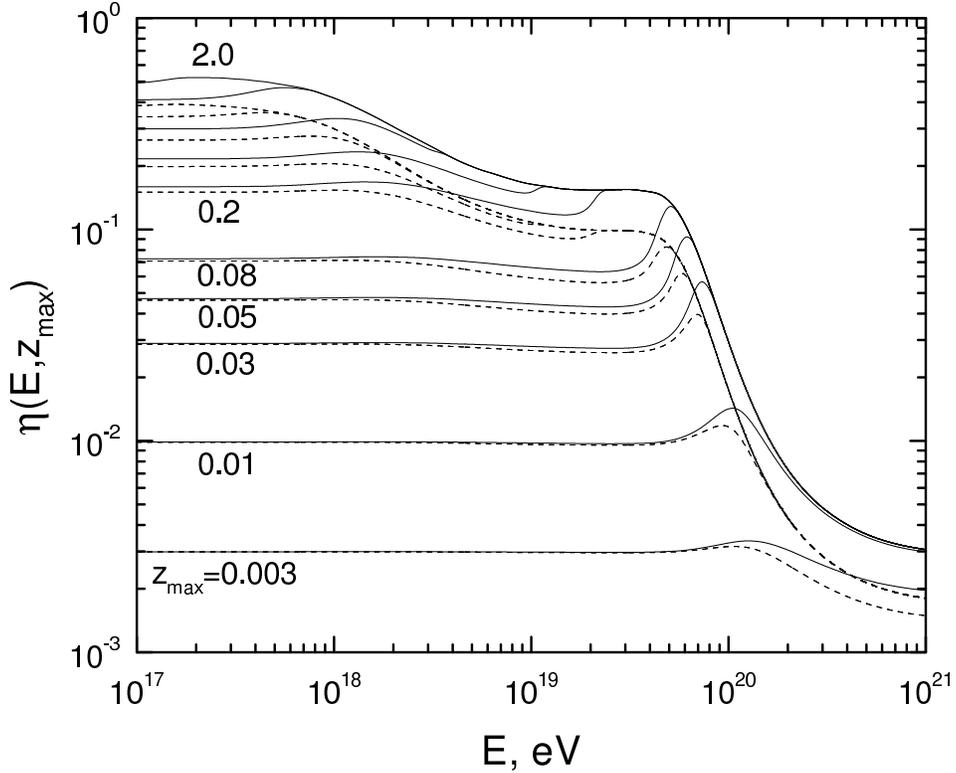}
  \end{center}
 \caption{Modification factors for diffuse spectra with different
    $z_{\rm max}$. The curves between $z_{\rm max}=2.0$ and    
$z_{\rm max}=0.2$ have $z_{\rm max}= 0.3~~,0.5~ $ and ~~ 1.0. The
    solid curves are for $\gamma_g=2.0$ and dashed ones - for 
 $\gamma_g=2.7$. The pile-up peaks, 
clearly seen at small $z_{\rm max}$,  disappear when summation of
    sources goes to large $z_{\rm max}> 1$.} 
  \label{bumps}
 \end{figure}
In Fig.\ref{bumps} the evolution of modification factor with growth of 
$z_{\rm max}$ is shown. The pile-up peaks are seen at small $z_{\rm max}$.
For large $z_{\rm max}$ they are located at different
energies, and their sum is given by a smooth curve without visible
bump(s). The solid curves correspond to $\gamma_g=2.0$ and the dashed ones
- to $\gamma_g=2.7$. 
In Fig.\ref{modif} we present two 
modification factors: $\eta_{ee}$ (dotted curve), when the adiabatic
energy losses and that due to pair-production are included, and 
$\eta_{\rm tot}$ (solid and dashed curves), when all energy losses are 
included. Both modification factors are normalized at low energies 
by values $\eta(z_{\rm max})$, given by Eq.({\ref{eta-lim}), 
so that their low-energy limit is equal
1. In case of $\eta_{ee}$ (dotted curve), it tends to 1 for both high and 
low energies, because pair-production energy losses disappear there. 
This curve is shown for the case $\gamma_g=2.7$.

Transition to the photopion energy losses is accompanied by practically
invisible bump.

If one displays the spectrum without factor $E^3$, it also shows no
presence of the bump.
\begin{figure}[ht]
 \begin{center}
    \includegraphics[height=20pc]{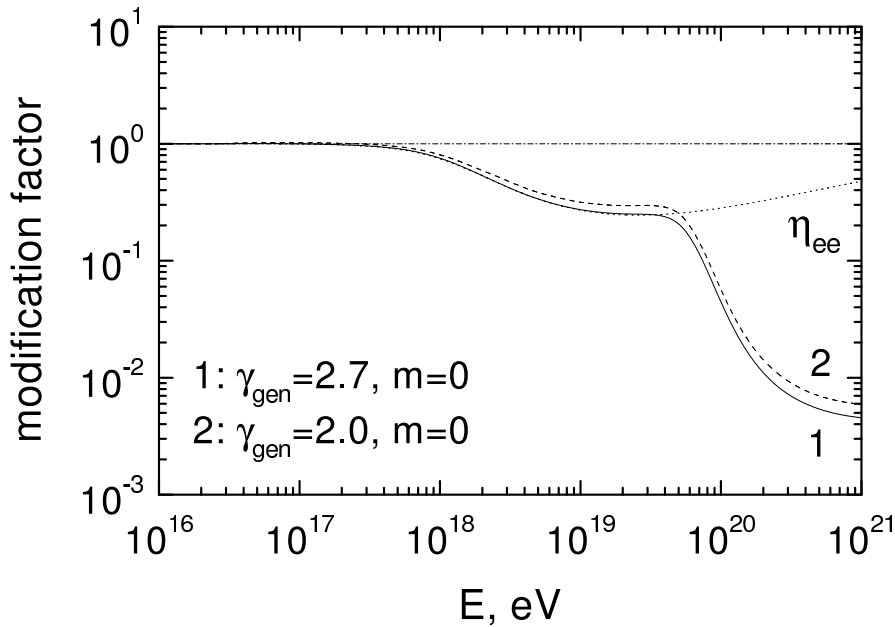}
  \end{center}
 \caption{Modification factor as characteristic of the dip and
bump. The dotted curve show $\eta_{ee}$, when adiabatic and pair-production 
energy losses are included, for the case $\gamma_g=2.7$. The solid 
and dashed curves include also the pion-production energy losses. 
The pile-up peaks are practically absent. } 
  \label{modif}
 \end{figure}
\section{AGN as UHECR sources}
AGN are traditional candidates for UHECR sources. The particles can be
accelerated there up to $E_{\rm max} \sim 10^{21}$~eV (Biermann and
Streitmatter 1987, Ipp and Axford 1991, Rachen and Biermann 1993). The
CR emissivity ${\cal L}_0 \sim 3\times 10^{46}$~erg/Mpc$^3$yr is well within
total emissivity of AGN, e.g. that of Seyfert galaxies is of order 
${\cal L}_{\rm Sy} \sim n_{\rm Sy}L_{\rm Sy} \sim 1\times 10^{48}$
~erg/Mpc$^3$yr.
AGN origin of UHECR results in presence of the GZK cutoff, and this
prediction agrees with data of Fly's Eye, HiRes and Yakutsk detector. 
In case AGASA spectrum is the correct one, the extra component at 
$E \geq 1\times 10^{20}$~eV is needed. In Fig.\ref{spectra} we plot in
the AGASA panel the spectrum of UHE photons from Superheavy Dark
Matter (SHDM), which is according to recent calculations 
(Berezinsky and Kachelriess 2000, Sarkar and Toldra 2001, 
Barbot and Drees 2002) is $I_{\gamma}(E) \propto E^{-2}$. Note that in
this case at energies $E< 1\times 10^{20}$~eV the protons dominate.  

The remarkable direct evidence for AGN as UHECR sources at energy 
range $(2 - 8)\times 10^{19}$~eV has been found in series of papers by 
Tinyakov and Tkachev 2001,2003 and references therein. 
There is the correlation between arrival directions of UHE particles 
in the AGASA and Yakutsk detectors and directions to BL Lacs, which are
AGN with jets directed towards us. This correlation implies that UHE signal 
carriers propagate rectilinearly in the universe. The rectilinear propagation
of the signal carriers from the point-like sources gives also the most natural 
interpretation of the small-angle clustering (Dubovsky, Tinyakov, Tkachev 2000;
Fodor and Katz 2000) observed by AGASA (Uchihori et al 2000). 

If UHECR are protons, as we argue above, the extragalactic magnetic fields
should be very weak. 

\section{Extragalactic magnetic fields and rectilinear propagation of
UHE protons}

How weak the magnetic field must be ?

Magnetic field must not produce the angular deflection larger than 
angular resolution of sources in the detectors, which is typically 
$\theta_{\rm res}\approx 2.5^{\circ}$. The correlation is found in
the energy range $(4 - 8)\times 10^{19}$~eV, for which the largest
attenuation length is $l_{\rm att} \sim 1000$~Mpc. The required 
upper limit for the     
magnetic field, which is homogeneous on this scale, is 
$B_l \leq 2\times 10^{-12}l_{1000}^{-1}$~G, where $l_{1000}$ is
attenuation length for $4\times 10^{19}$~eV protons in units of 1000 Mpc.
For a magnetic field with small homogeneity length,  $l_{\rm hom}$, the 
required upper limit is 
\begin{equation} 
B \leq \frac{E\theta_{\rm res}}{e\sqrt{l_{\rm att}l_{\rm hom}}} \sim
6\times 10^{-10}~\mbox{G},
\label{B}
\end{equation}
where the numerical value is given for $l_{\rm att} \sim 1000$~Mpc 
and $l_{\hom} \sim 10$~kpc.

We argue that these fields are not excluded. 

The observed Faraday rotations give only the upper limits on large
scale extragalactic magnetic field (Kronberg 1994).
All known mechanisms of generation of the large scale cosmological 
magnetic field results in extremely weak magnetic field 
$\sim 10^{-17}$~G or less (for a review see Grasso, Rubinstein 2001). 
The strong magnetic field can be generated in compact 
sources, 
presumably by dynamo mechanism, and spread away by the flow of the
gas. These objects thus are surrounded by magnetic halos, where
magnetic field can be estimated or measured. 
The strong magnetic fields of order of $1\mu$G are indeed
observed in galaxies and their halos, in clusters of galaxies and 
in radiolobes of radiogalaxies. As an example one can consider our
local surroundings. Milky Way belongs to the Local Group (LG)
entering the Local Supercluster (LS). LG with a size 
$\sim 1$Mpc contains 40 dwarf galaxies, two giant spirals (M31
and Milky Way) and two intermediate size galaxies. The galactic winds
cannot provide the appreciable magnetic field inside this structure. 
LS with a size of 10 -- 30 Mpc is a young system 
where dynamo mechanism cannot amplify significantly the  primordial 
magnetic field. 
In fact LS is filled by galactic clouds submerged in the voids. 
The vast majority of the
luminous galaxies reside in a small number of clouds: 98 \% of all
galaxies in 11 clouds (Tully 1982). Thus, accepting the hypothesis
of generation of magnetic fields in compact sources, one arrives at 
the perforated  picture of the universe, with strong magnetic fields in the
compact objects and their halos (magnetic bubbles produced by galactic 
winds) and with extremely weak magnetic fields outside.  However,
even in this picture there is a scattering of UHE protons off the
magnetic bubbles and the  scattering length is 
$l_{\rm sc} \sim 1/\pi R^2 n$, where $R$ is the radius of magnetic bubbles
and $n$ is their space density. Among various 
 structures, the largest contribution is given by galaxy clusters which
can provide $l_{\rm sc} \sim (1-2)\times 10^3$~Mpc.  

\section{Discussion}
In this paper we consider AGN as UHECR sources. We assume that
there are two mechanisms of acceleration operating there: one accelerates
protons along the jet (Chen et al 2002), while the other operates in
the shock where the jet terminates (Biermann and Streitmatter 1987,  
Ipp and Axford 1991, Rachen and Biermann 1993). When a jet is directed
towards us, we observe the jet accelerated protons and classify the AGN
as BL Lac. The shock-accelerated protons are emitted isotropically,
and their flux at the earth is smaller than that from the jet
component because of
the geometrical factor. We assume additionally that CR luminosity of the 
jet component is larger than of isotropic one. Then BL Lacs are to be
observed as UHECR sources at large distances. At small distances, less 
than 135 Mpc (attenuation length of proton with energy 
$1\times 10^{20}$~eV), the total number of AGN is smaller, and the
probability to observe an AGN as BL Lac is correspondingly smaller. 
But AGN can be detected at these distances due to isotropic
component. At the distances less 31.5 Mpc (attenuation length of a
proton with energy $2\times 10^{20}$~eV) there are as minimum 11 Seyfert
galaxies (with redshifts $z\leq 0.009$) and 4 nearby radiogalaxies. 
The lower limit on the number of UHECR sources for this distance is
estimates by Tinyakov and Tkachev 2003 on basis of small-angle
clustering  as 30 at 90\% CL and 4 at 
99\% CL, in agreement with observed number of AGN given above. For 
135 Mpc ($E_p=1\times 10^{20}$~eV the number of sources become
80 times larger. It justifies the use of uniform distribution of the sources
for flux calculations at $E\leq 1\times 10^{20}$~eV. 

For events at energies $E> 1\times 10^{20}$~eV the AGN model meets the 
difficulties due to short attenuation length of protons (31.5 Mpc for 
$E_p=2\times 10^{20}$~eV). UHECR sources must be seen in the direction of
each proton with such energy. One can consider two cases.\\
\noindent
(i) The AGASA data at $E\geq 1\times 10^{20}$~eV are included in
analysis. The AGASA
excess should be explained by another component as shown in 
Fig.\ref{spectra}\\
\noindent
(ii) If to neglect AGASA data, one has to analyse 
only three events with energies $E\geq 1\times 10^{20}$~eV, namely
one Fly's Eye event
with $E \approx 3\times 10^{20}$~eV, one HiRes event with 
$E \approx 1.8\times 10^{20}$~eV, and one Yakutsk array with      
$E \approx 1.0\times 10^{20}$~eV. Let us consider the highest energy 
Fly's Eye event. If it is induced 
by a proton and its source is an AGN, then the distance to it should
be only 20 - 30 Mpc. No AGN is observed in this direction and
at this distance. As in case (i) we again have to assume the new component at 
$E \geq 1\times 10^{20}$ eV, but this time it is based on very low 
statistics.

To avoid the problems with $E\geq 1\times 10^{20}$~eV within the AGN
model, one can assume the presence of 
strong extragalactic magnetic field. The AGN model in this case  
explains all data, except the AGASA excess and correlation with BL Lacs. 
In particular the small-angle clustering can be
explained by magnetic lensing (Harari et al 1999, Sigl et al 1999,  
Yoshiguchi et al 2002). One can try to explain the correlation with 
BL Lacs by flux of light neutral hadrons from AGN (Berezinsky et al 2002, 
Kachelriess et al 2003). However, in the case we want to explain the dip 
observed in UHECR by primary protons, the flux of light hadrons produced in 
AGN must be {\em subdominant} with more flat spectrum ($\propto E^{-2.7}$)
than protons. As a result the correlation must increase with energy, which
is not observed. The number of correlated events is expected to be
small, being 
proportional to the fraction of events induced by light neutral hadrons.  

\section{Conclusions}

The preliminary data of HiRes indicate to proton composition of UHECR
at $E \geq 1\times 10^{18}$~eV (Sokolsky 2002).
The observed energy spectra reveal 
the signatures of interaction of UHE protons with the CMB in the form of
the dip, beginning of the GZK cutoff and the good agreement with 
the predicted spectrum. Combined with small-angle clustering (Uchihori
et al 2000)
and correlation with BL Lacs (Tinyakov and Tkachev 2001, 2003), 
these data require the 
rectilinear propagation of UHE protons. 
{\em The correlation with AGN (BL Lacs) becomes thus the most sensitive method
of measuring extragalactic magnetic fields, in case they are very weak}. 

Events with $E\geq 1\times 10^{20}$~eV imply the new
component, which can be e.g. gamma-rays from the decay of superheavy 
particles. This case is especially favorable for 12 AGASA events with
theses energies. The spectrum  of UHE photons from the decay of 
superheavy particles fits well the AGASA observations 
(see Fig.\ref{spectra}).

\section{Acknowledgment}
One of the authors (V.B.) is grateful to the organizers of the
Workshop, and in particular to Prof. M.Teshima for kind invitation.
We are grateful to Pierre Sokolsky for active discussion and for providing
us with HiRes data. Many thanks to Masahiro Teshima for the data of AGASA,
and to M.I.Pravdin for the data of Yakutsk.
We gratefully acknowledge the useful discussions with Peter Biermann, 
Zurab Berezhiani, Michael Kachelriess, Karl-Heinz Kampert and Igor Tkachev.  
D.R.Bergman is thanked for information concerning the HiRes data.

This work was supported in part by INTAS (grant No. 99-01065).

\section{References}
\noindent
\ Abu-Zayyad T. et al (HiRes collaboration)\ 2002a, astro-ph/0208243 
\re
\ Abu-Zayyad T. et al (HiRes collaboration)\ 2002b, astro-ph/0208243
\re
\ Barbot C. and Drees M.,\ 2002, Phys. Lett. B 533, 107 
\re
\ Berezinsky V.S. and Grigorieva S.I.,\ 1988, Astron.Astroph. 199, 1.
\re
\ Berezinsky V., Gazizov A.Z., and Grigorieva S.I.,\ 2002 hep-ph/0204357.
\re
\ Berezinsky V. and Kachelriess M.,\ 2001, Phys. Rev. D 63, 034007
\re
\ Berezinsky V., Kachelriess M., and Ostapchenko S.\ 2002,
Phys. Rev. D 65, 083004
\re
\ Biermann P.L. and Streitmatter P.A.,\ 1987, Ap. J., 322, 643
\re
\ Biermann P.L. et al,\ 2003, astro-ph/0302201
\re 
\ Bird D.J. et al (Fly's Eye collaboration)\ 1994, Ap. J., 424, 491
\re
\ Chen P., Tajima T., and Takahashi Y., \ 2002, Phys. Rev. Lett. 89, 161101
\re
\ Dubovsky S.L., Tinyakov P.G. and Tkachev I.I.,\ 2000, Phys. Rev. Lett. 
85, 1154 
\  Fodor Z. and Katz S.D.,\ 2000, Phys. Rev. D63, 023002 
\re
\ Glushkov A.V. et al (Yakutsk collaboration)\ 2000, JETP Lett. 71, 97 
\re
\ Glushkov A.V. and Pravdin M.I.\ 2001, JETP Lett. 73, 115
\re
\ Grasso D., Rubinstein H.,\ 2001, Phys. Rep. 348, 163 
\re
\ Greisen K.,\ 1966, Phys. Rev. Lett., 16, 748 
\re
\ Harari D., Mollerach S., and Roulet E., \ 1999, JHEP 08, 022
\re
\ Hayashida N. et al,\ 1999, Astrop. Phys. 10, 303
\re
\ Hill C.T. and Schramm D.N.,\ 1985, Phys. Rev. D 31, 564
\re
\ Hoerandel J.R.,\ 2002, astro-ph/0210453, to be published in 
Astrop. Phys.
\re
\ Ipp W.H. and Axford W.I.,\ 1991, Astrophysical Aspects of Most Energetic
Cosmic Rays (ed. M. Nagano), World Scientific, 273 
\re
\ Kachelriess M., Semikoz D.V. and Tortola M.A.,\ 2003, hep-ph/0302161
\re
\ Kampert K-H.,\ 2001, astro-ph/0101133, Symposium on Fundamental
Issues in Elementary Matter, Bad Honnef, Germany
\ Kronberg P.P.,\ 1994, Rep. Prog. Phys. 57, 325
\re
\ Rachen J.P. and Biermann P.L.,\ 1993, Astron. Astroph.  272, 161 
\re
\ Sarkar S. and Toldra R.,\ 2002, Nucl. Phys. B 621, 495
\re
\ Sigl G., Leimone M., and Biermann P., \ 1999, Astrop. Phys. 10, 141 
\re
\ Sokolsky P.,\ 2002, ``The High Resolution Fly's Eye - Status and Preliminary
Results on Cosmic Ray Composition above $10^{18}$~eV'',  
Proc. of SPIE Conf on Instrumentation for Particle Astrophysics, Hawaii
\re
\ Stanev T., Engel R., Muecke A., Protheroe R.J. and Rachen J.P.,\ 2000,
Phys. Rev. D 62, 093005
\re
\ Takeda M. et al (AGASA collaboration),\ 2002, astro-ph/0209422
\re 
\ Tinyakov P.G. and Tkachev I.I.,\ 2001, JETP Lett., 74, 445
\re
\ Tinyakov P.G. and Tkachev I.I.,\ 2003, astro-ph/0301336
\re
\ Tully R.B.,\ 1982,  Ap. J.,  257, 389 
\re
\ Uchihori Y., et al,\ 2000, Astrop. Phys.  13, 157 
\re
\ Yoshiguchi H., Nagataki S., Tsubaki S., and Sato K., \ 2002, 
astro-ph/0210132
\re
\ Yoshida S., and Teshima M.,\ 1998, Progr. Theor. Phys. 89, 833
\re
\ Zatsepin G.T. and Kuzmin V.A.,\ 1966, Pisma Zh. Experim. Theor. Phys.,
4, 114
\re

\endofpaper 
\end{document}